\documentclass[journal=jacsat,manuscript=article]{achemso}

\usepackage{bm}
\usepackage{amssymb}
\usepackage{graphicx}
\usepackage{amsmath}
\usepackage{textcomp}
\usepackage{colordvi}
\usepackage{multirow}
\usepackage{ulem}
\usepackage{tikz}
\usepackage{color}
\usepackage{makecell}
\usepackage{booktabs}
\usepackage{braket}

\usepackage{chemformula} 
\usepackage[T1]{fontenc} 

\author{Kainan Chang}
\email{knchang@ciomp.ac.cn}
\affiliation{GPL Photonics Laboratory, State Key Laboratory of Luminescence Science and Technology, Changchun Institute of Optics, Fine Mechanics and Physics, Chinese Academy of Sciences, Changchun 130033, China.
}

\author{Ying Song}
\affiliation{State Key Laboratory of Advanced Manufacturing for Optical Systems, Changchun Institute of Optics, Fine Mechanics and Physics, Chinese Academy of Sciences, Changchun 130033, China.
}

\author{Yuwei Shan}
\affiliation{GPL Photonics Laboratory, State Key Laboratory of Luminescence Science and Technology, Changchun Institute of Optics, Fine Mechanics and Physics, Chinese Academy of Sciences, Changchun 130033, China.
}

\author{Jin Luo Cheng}
\email{jlcheng@ciomp.ac.cn}
\affiliation{GPL Photonics Laboratory, State Key Laboratory of Luminescence Science and Technology, Changchun Institute of Optics, Fine Mechanics and Physics, Chinese Academy of Sciences, Changchun 130033, China.
}

\title{Harmonic generation of graphene quantum dots in Hartree-Fock approximation}


\begin{document}

\begin{tocentry}



\includegraphics[scale=0.4]{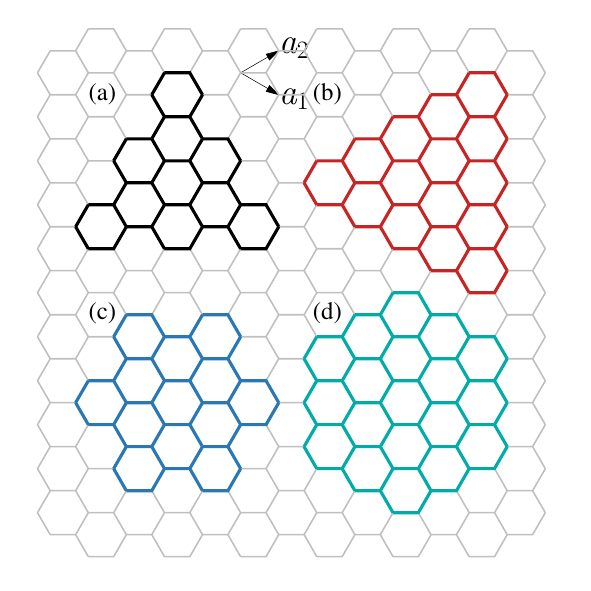} 



\end{tocentry}

\begin{abstract}

We theoretically investigate harmonic generation in graphene quantum dots under linearly polarized optical pulses, focusing on excitonic effects.
Combining the tight-binding model and  the single-particle density matrix approach, we derive semiconductor Bloch equations under a static-screened Hartree-Fock approximation. 
This framework characterizes the electron-electron interaction through local Hartree potentials for direct Coulomb interaction and nonlocal Fock potentials for exchange interaction. 
Distinct confgurations of Hartree and Fock terms yield various approximation methods, including the independent-particle approximation, the mean-feld approximation, the random phase approximation, and the excitonic effects. 
We thoroughly analyze how these approximation methods affect the electronic energy levels, linear optical absorption, and nonlinear harmonic generation. 
Within excitonic effects, we 
present the dependence of harmonic generation on the geometric variations of graphene quantum dots 
(sizes, triangular/hexagonal shapes, and armchair/zigzag edges) and the amplitude and polarization of electric fields.
Our findings show that excitonic effects significantly enhance optical responses of graphene nanostructures.
For a dot ensemble formed by randomly oriented graphene quantum dots, only odd-order harmonics exist along the polarization direction of the incident light.
Crucially, harmonic generation in graphene quantum dots exhibits  high tunability via geometric configuration,
making them promising candidates for nonlinear optical nanodevices.

\end{abstract}



Graphene, as the first fabricated two-dimensional material, has attracted enormous attention due to its chemical stability, mechanical robustness, high carrier mobility \cite{Tan2020,Brida2013},  ultrafast carrier dynamics \cite{Tokman2019, Oum2014}, wideband absorption from THz to visible \cite{Autere2018,Tan2020, Jiang2018, Mikhailov2007},  and extremely strong optical nonlinearity \cite{Mikhailov2008,Zhou2022}.
These properties arise from its linear dispersion and can be well controlled by Fermi levels through electric gating or chemical doping techniques \cite{Bonaccorso2010,Wang2008,Tomadin2018}. 
Given the easy integration with photonic structures, graphene is ideal for realizing photonic devices with novel functionalities.


However, applications in nano-electronic and nano-photonic devices are also limited by the zero bandgap \cite{Yavari2010,Cai2014},
which can be conquered by the advancements in nano-fabrication techniques \cite{Datta2008,Campos2009,Li2022,Guelue2010} via engineering the size, shape, or edge of graphene on the nanoscale to form various graphene nanostructures, including nanoribbons \cite{CastroNeto2009,Dutta2010,Wei2023}, flakes \cite{Sadeq2018,Aguillon2023} and quantum dots \cite{Sk2014,Tiutiunnyk2023}.
In particular, graphene quantum dots (GQDs) are zero-dimensional fragments of graphene with a size smaller than 100 nm, and have been widely applied in photovoltaics, light-emitting diodes, batteries, fuel cells, memory devices, sensitisers in solar cells, bioimaging, etc. \cite{Buzaglo2016,Sk2014,Sadeq2018,Nesakumar2022,Kalluri2023}.

Recent studies have focused on understanding the nonlinear optical properties of GQDs.
Cox {\it et al.} found that the second- and third-order polarizabilities of GQD exceed those of noble metal nanoparticles with similar lateral size  \cite{Cox2014}, and the physical mechanism is attributed to the strong plasmonic near-field enhancement; its polarization can be precisely tuned and controlled by applying an external electric field   \cite{Cox2017}.
Aguillon {\it et al.} \cite{Aguillon2023} theoretically investigated the effect of vacancy defects on the second harmonic generation (SHG) of GQD, and obtained an efficient SHG even for vacancy concentrations as low as one per several thousands of lattice atoms.
Avchyan {\it et al.} \cite{Avchyan2022} and Gnawali {\it et al.} \cite{Gnawali2022,Gnawali2023} showed the dependence of harmonic generation on the size, edge, and shape of GQD.
Among these works, the Coulomb interaction is considered either at the level of mean-field  approximation (MFA) or random-phase approximation (RPA), while excitonic effects (EXE) are not taken into account.
However, it is well known that EXE in two-dimensional materials with a bandgap are extremely important for both linear \cite{Guelue2010,Yang2011} and nonlinear \cite{Mueller2018,Song2023} optical properties due to the insufficient screening of the Coulomb interaction,  and it is of fundamental importance to understand how it affects nonlinear optical responses in GQDs.

\begin{figure}[htb]
\centering
\includegraphics[scale=0.9]{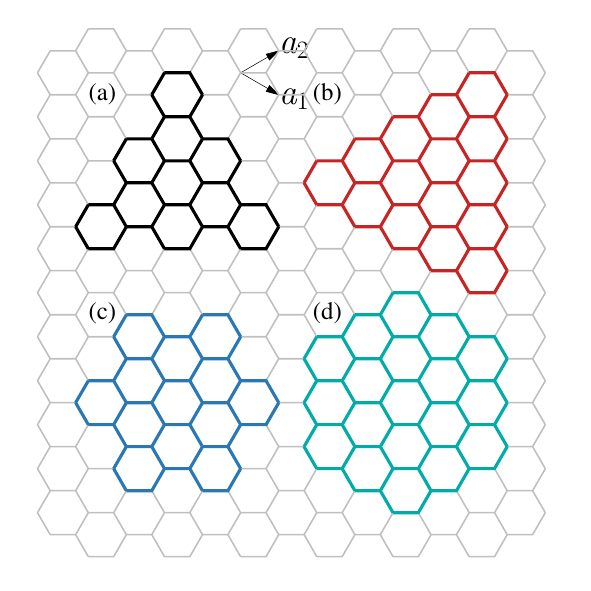} 
\caption{Illustration of GQDs with different edges and shapes: (a) 3-AC-TQD
  (b) 5-ZZ-TQD, (c) 2-AC-HQD, and (d) 3-ZZ-HQD. $\bm a_1$ and $\bm a_2$ indicate
  the primitive lattice vectors of graphene. 
  }
\label{fig-1}
\end{figure}

For this purpose,
we focus on the harmonic generation in GQDs, 
with different shapes and edges as shown in Fig.\,\ref{fig-1}.
The electronic states of GQDs are calculated in the framework of tight-binding approximation with the nearest-neighbor coupling, and Coulomb electron-electron interactions at the level of static-screened Hartree-Fock (HF) approximation.
The dynamics of applying a strong laser field is described by semiconductor Bloch equations,
and their numerical solutions are used to extract the nonlinear susceptibilities.
Within our model, the Hartree term of the Coulomb interaction contributes to the local field and RPA, while the Fock term gives rise to EXE.
Our results reveal that the EXE has a profound impact on the resonance energy and intensity of the optical responses of GQDs. 
Furthermore, we also discuss the effects of the size, edge, and shape of GQDs, as well as the amplitude and polarization of the electric field.

%


\section{Results and discussion}
\label{results}

We consider the GQDs, which are nanoflakes with different edges and shapes cut from a monolayer graphene, 
including equilateral triangular QDs with armchair edges (AC-TQD) or zigzag edges (ZZ-TQD)
and hexagonal QDs with armchair edges (AC-HQD) or zigzag edges
(ZZ-HQD), as shown in Figs.\,\ref{fig-1}\,(a--d), respectively. These GQDs are described by the
number of hexagons
at each edge as $N$-AC-TQD, $N$-ZZ-TQD,
$N$-AC-HQD, and $N$-ZZ-HQD.

\subsection{Electronic structures at screened HF approximation}

\begin{figure*}[htb]
\centering
\includegraphics[scale=0.7,angle=90]{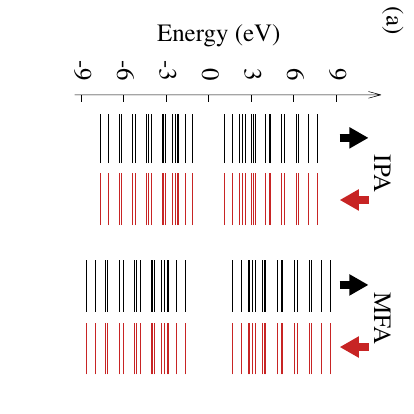} 
\includegraphics[scale=0.7,angle=90]{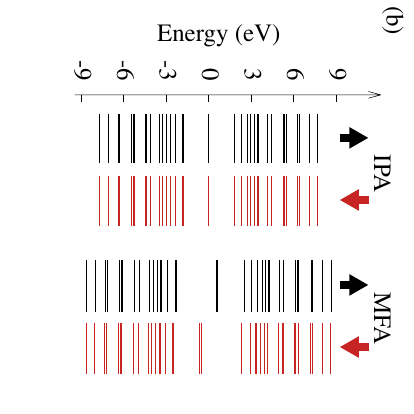} 
\includegraphics[scale=0.7]{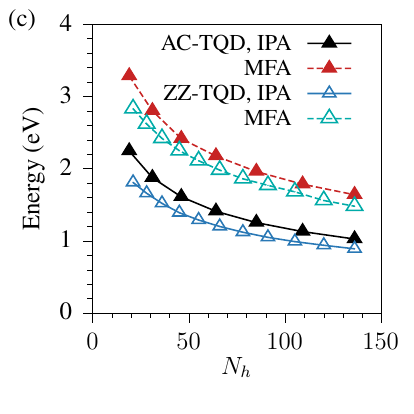} \\
\includegraphics[scale=0.7,angle=90]{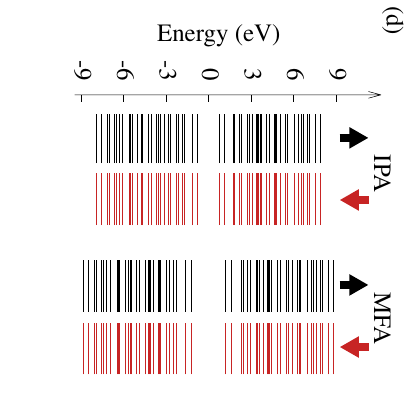} 
\includegraphics[scale=0.7,angle=90]{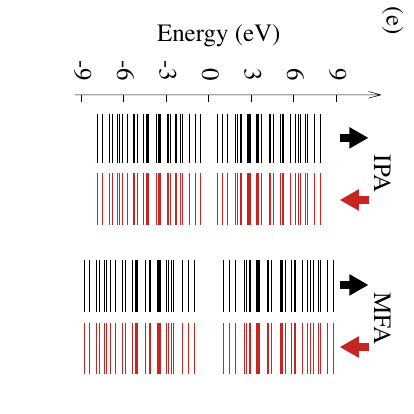} 
\includegraphics[scale=0.7]{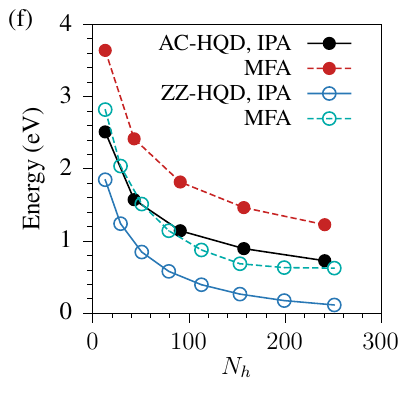} 
\caption{Electronic energy levels at IPA (left panel) and MFA (right panel), for (a) 4-AC-TQD, (b) 6-ZZ-TQD, (d) 3-AC-HQD, (e) 4-ZZ-HQD.
Colored arrows mean different spins. 
Size-dependent bandgap for (c) TQD and (f) HQD.
}
\label{fig-2}
\end{figure*}

We first discuss how the static-screened HF approximation
  affects the energy levels of GQD, mainly focusing on the
  eigenenergies $\varepsilon_{s\sigma}$ from Eq.\,\eqref{eq:sch}. From
  our calculations,  the ground states of all studied GQDs can be taken
  as antiferromagnetic states, which have lower energies. 
Figures\,\ref{fig-2}\,(a), (b), (d), and (e) show energy levels of
4-AC-TQD, 6-ZZ-TQD, 3-AC-HQD, and 4-ZZ-HQD in
independent-particle
  approximation (IPA) and MFA, respectively.  Because
   only the nearest neighbor coupling is considered in the tight-binding
   model, the electron-hole symmetry is preserved for all considered GQDs
   in IPA, like that for graphene. Due to the high symmetry of these GQDs, there exist many degeneracies in IPA electronic
   states, as reported by G\"{u}\c{c}l\"{u} {\it et al.} \cite{Guelue2010}.
   Except for the 6-ZZ-TQD, there exists an energy gap between the lowest unoccupied molecular orbital  (LUMO) and the highest occupied molecular orbital (HOMO) states; while for 6-ZZ-TQD, there 
   exist 10 degenerate states with zero energy (with the inclusion of spin degeneracy)
   but only half of them are occupied,  which leads to zero gaps.
By taking into account the screened HF approximation in MFA, the electronic states for all GQDs are significantly changed. For ZZ-TQD, antiferromagnetic ground state is more stable,
   and the degeneracy between spin states is broken; however, each
   spin up state with energy $\epsilon$ is accompanied by a spin down
   state with energy $-\epsilon$. At the same time, all zero-energy
   states are split into two spin groups with different energies, and
   each group includes five energy levels with very close energies, which
   gives a nonzero gap. For other types of GQDs, the total energies for 
   antiferromagnetic ground state and paramagnetic ground state are
   very close to each other, but the former is slightly lower; the
   spin degeneracy and electron-hole symmetry are both preserved, but
   the energy gaps are widened, which is as expected because the screened HF
   approximation gives the lowest order of GW correction. 
Figures\,\ref{fig-2}\,(c) and (f) give the dependence of the
  bandgap on the QD size.  In general, the gaps in MFA are larger than those in IPA, indicating the important effects
  of Coulomb interaction. All gaps decrease with increasing size of GQDs,
  because of weaker quantum confinement.


\subsection{Linear optical response}

\begin{figure*}[!htb]
\centering
         \includegraphics[scale=0.85]{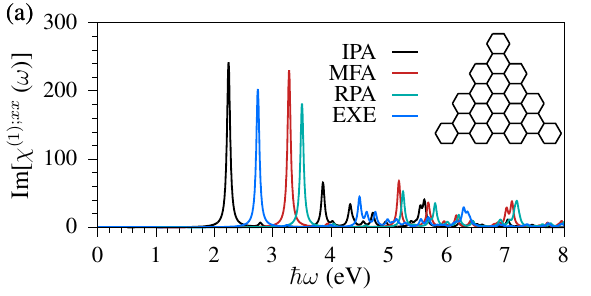}
         \includegraphics[scale=0.85]{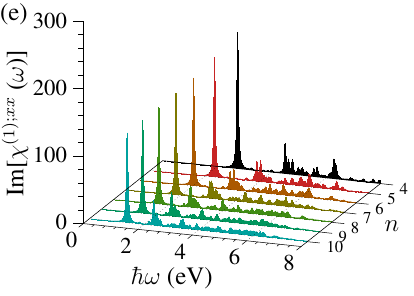} \\
         \includegraphics[scale=0.85]{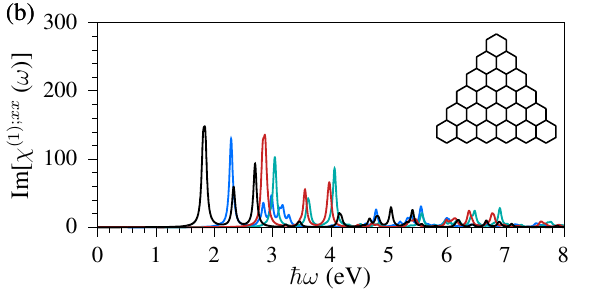}
         \includegraphics[scale=0.85]{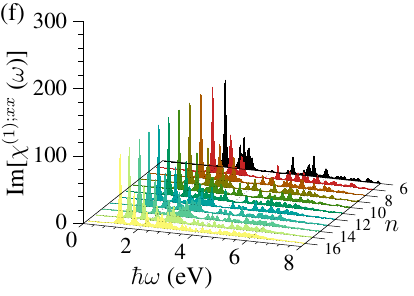} \\
         \includegraphics[scale=0.85]{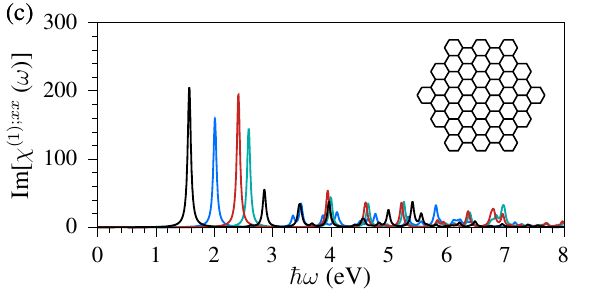}
         \includegraphics[scale=0.85]{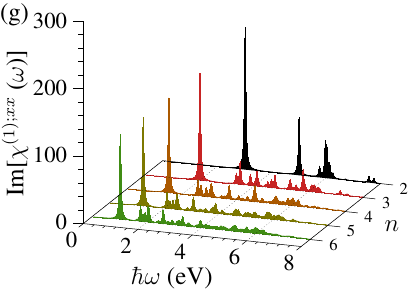} \\
         \includegraphics[scale=0.85]{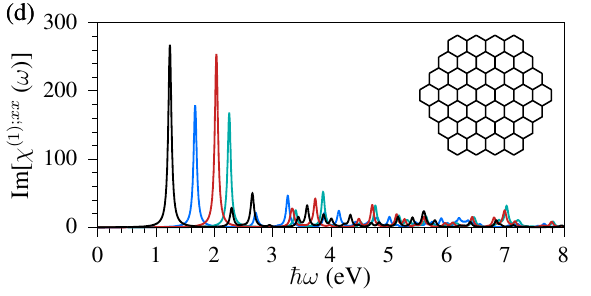}
         \includegraphics[scale=0.85]{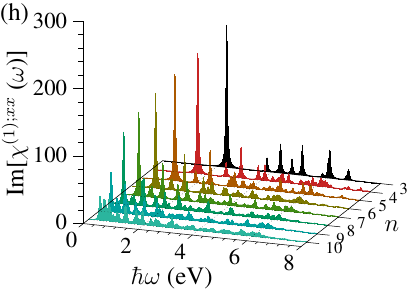} \\
     \caption{Spectra of  $\text{Im}[\chi^{(1);xx}(\omega)]$ for (a)  4-AC-TQD, (b)
   6-ZZ-TQD, (c) 3-AC-HQD, (d) 4-ZZ-HQD under different approximations of IPA, MFA, RPA, and EXE. 
    (e--h) QD size dependence of the spectra in EXE for the shapes as those in (a--d).}
     \label{fig-3}
\end{figure*}

We consider the optical response of the polarization to an applied electric field pulse with a Gaussian envelope function,
  \begin{align}
    \bm E(t) = E_0 \hat{\bm e}_\Omega e^{-t^2/\Delta_c^2} e^{-i\Omega t}+\rm c.c.\,,
    \end{align}
with the duration $\Delta_c$, the center frequency $\Omega$, and the polarization $\hat{\bm e}_{\Omega}$. In the
frequency domain, this corresponds to a function with Gaussian peaks
as
\begin{align}
  \bm E(\omega) = &\int dt \bm E(t) e^{i\omega t} \notag\\
  =& \sqrt{\pi}\Delta_c
  \left[\hat{\bm e}_\Omega e^{-(\omega-\Omega)^2\Delta_c^2/4}  +
  \hat{\bm e}_\Omega^\ast e^{-(\omega+\Omega)^2\Delta_c^2/4}\right]\,.
\end{align}
The polarization $\bm P(t)$ is also a pulse, and its spectra $\bm
P(\omega)=\int dt \bm P(t) e^{i\omega t}$ are located around the
harmonic frequencies $j\Omega$  for integer $j$. 
A linear susceptibility tensor is extracted from
\begin{align}
\chi^{(1);da}(\omega)=\frac{P^d(\omega)}{E^a(\omega)}\,.
\end{align}

In the following calculations, we take $T=300$ K, $\mu=0$ eV, and $\Gamma=20$ fs.
For the linear optical process,
a simple numerical method to extract the susceptibility spectra is to perform the excitation by a very short laser pulse with the parameters $\Delta_c=1 $ fs, $\hbar\Omega=3$ eV, and $E_0=10^4$ V/m. 
For all considered geometries, the symmetries determine the nonzero susceptibility components as $\chi^{(1);xx}=\chi^{(1);yy}$. 
The calculated absorption spectra, given by $\text{Im}[\chi^{(1);xx}(\omega)]$, are plotted in Figs.\,\ref{fig-3}\,(a--d)
for the GQDs in Figs.\,\ref{fig-2}\,(a, b, d, e), under different approximations of IPA, MFA, RPA, and EXE.
Compared with the spectra in IPA, the peaks of those in MFA move to higher energies. 
It is because that under the single-particle approximation, the optical absorption processes are directly determined by the energy levels shown in Fig.\,\ref{fig-2}, and the lowest-energy peaks are dictated by the energy gaps \cite{Cheng2019,Yang2009,Hwang2007}. 
When the Coulomb interaction is taken into account through the Hartree term in RPA, 
the absorption spectra blue-shift with weaker amplitudes, compared with the case in MFA.
The primary reason is that 
the total electric field experienced by the electrons in RPA is not only the external field $\bm E(t)$, but also the field induced by the inhomogeneous charge density generated under the optical excitation. 
Due to the geometric effect, these local plasmons can be excited by plane waves.
In this way, the absorption peak locations in RPA are shifted to higher energies corresponding to the local plasmon resonance.
When both Hartree and Fock terms are considered in EXE, the absorption spectra red-shift to lower the optical gap, which indicates the formation of excitonic levels.
It is also clear that the amplitudes of the first peaks in EXE, associated with the lowest exciton energy, are higher than those in RPA, indicating the enhancement of linear absorption by excitonic effects.

Figures\,\ref{fig-3}\,(e--h) show the spectra of $\text{Im}[\chi^{(1);xx}(\omega)]$ for GQDs with different sizes at the level of EXE.
The first peaks give a strong absorption at the lowest exciton energy.
As the QD size increases, these first peaks move to lower 
energies, because of the weaker quantum confinement energy for GQDs with larger size; and the amplitudes of spectra become lower, due to the weaker optical transitions.
%

\subsection{Nonlinear Optical Response}

Then we turn to the nonlinear optical responses, in which the electronic polarization $\bm P(\omega)$ is located around the center frequencies $j\Omega$ for $j>1$. Perturbatively, this polarization can be written as \cite{Cheng2015}
\begin{align} 
\int P^d(j\Omega+\delta) d\delta = \int & \frac{d\delta_1 d\delta_2\cdots \delta_j}{(2\pi)^j}
\sum_{a_1,a_2,\cdots,a_j}
\chi^{(j);da_1a_2\cdots a_j}(\Omega+\delta_1,\Omega+\delta_2,\cdots,\Omega+\delta_j) \notag\\
& E^{a_1}(\Omega+\delta_1)\cdots E^{a_j}(\Omega+\delta_j)\,.
  \label{eq:p}  
\end{align} 
Taking the approximation 
\begin{align} 
\sum_{a_1,a_2,\cdots,a_j}\chi^{(j)}(\Omega+\delta_1,\Omega+\delta_2,\cdots)[\hat{\bm e}_\Omega]^{a_1}[\hat{\bm e}_\Omega]^{a_2}\cdots \approx \overline{\chi}^{(j);d}(\Omega,\hat{\bm e}_\Omega)\,,
\end{align} 
the effective susceptibility
$\overline{\chi}^{(j);d}(\Omega,\hat{\bm e}_\Omega)$ can be calculated through
\begin{align}
  \overline{\chi}^{(j);d}(\Omega,\hat{\bm e}_\Omega) = 
  \frac{1}{E_0^j}\int_{-\Omega/2}^{\Omega/2} P^d(j\Omega +\delta) d\delta \,,\quad\text{for}\,(j\geq2)\,.
  \label{eq:chi}                       
\end{align}
Additionally,
in the following calculations of nonlinear responses, the pulse duration is fixed at $\Delta_c=30$ fs.

\subsubsection{Spectra of $\bm P(\omega)$ for $\hbar\Omega=0.91$ eV and $E_0=10^8$ V/m}

\begin{figure*}[!htb]
         \includegraphics[scale=0.8]{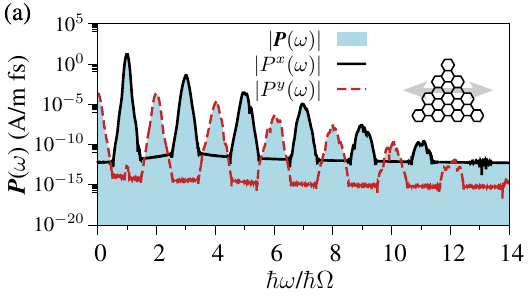}
         \includegraphics[scale=0.8]{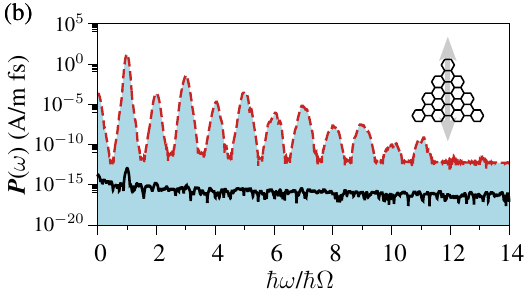}       
         \includegraphics[scale=0.8]{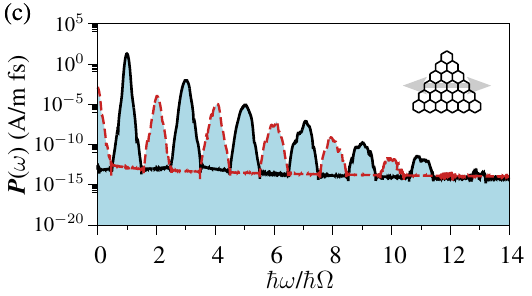}
         \includegraphics[scale=0.8]{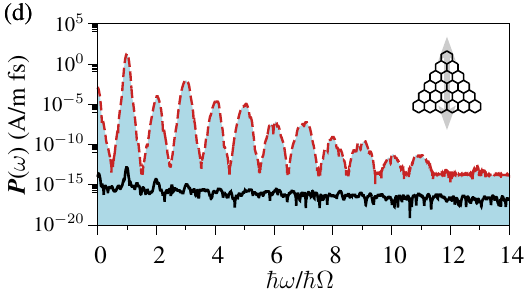}        
         \includegraphics[scale=0.8]{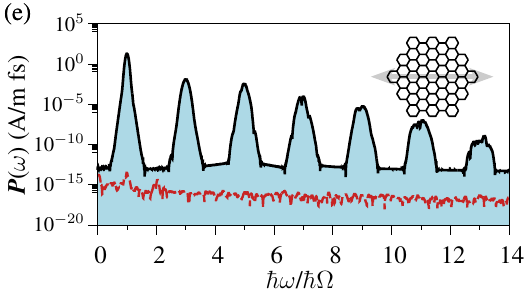}
         \includegraphics[scale=0.8]{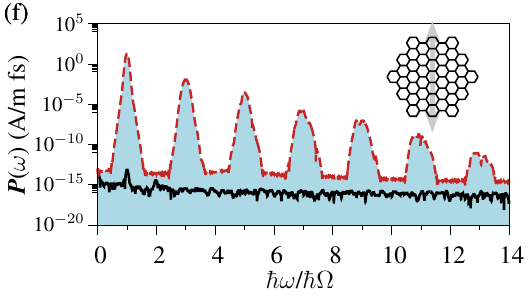}  
         \includegraphics[scale=0.8]{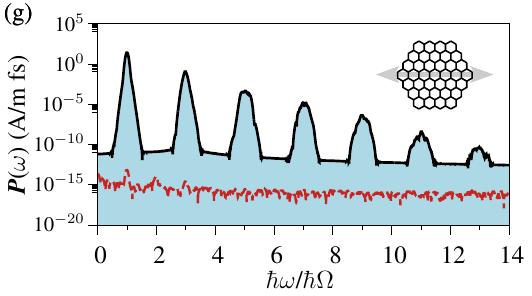} 
         \includegraphics[scale=0.8]{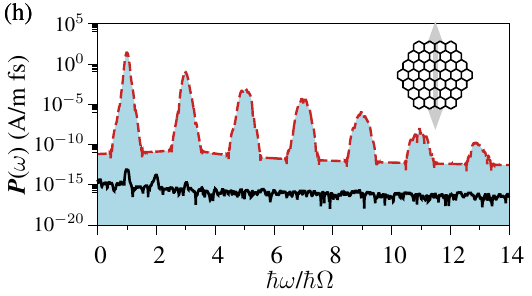}
      \caption{Spectra of $|{\bm P}(\omega)|$, $|P^x(\omega)|$, and $|P^y(\omega)|$ for different types of GQDs (a, b) 4-AC-TQD,  (c, d) 6-ZZ-TQD, (e, f) 3-AC-HQD, and (g, h) 4-ZZ-HQD. The light parameters are $\hbar\Omega=0.91$ eV, $E_0=10^8$ V/m, and the field direction is along the $x$-direction (a, c, e, g) or the $y$-direction (b, d, f, h), as shown in the insets.}
      \label{fig-4}
\end{figure*}

We first consider the field polarization dependence of the spectra  $\bm P(\omega)$ for different types of GQDs in the EXE approximation, and the results are shown in Fig.\,\ref{fig-4}, where the light parameters are $\hbar\Omega=0.91$ eV and $E_0=10^8$ V/m. 
It can be seen that all HQDs and TQDs can generate odd-order harmonics up to the thirteenth order, while only TQDs can generate even-order harmonics. 
For odd-order harmonics, which can be generated from all these four types of GQDs, their polarization is along the field polarization, either along the $x$ or $y$ direction with the response coefficients satisfying $\chi^{(2j+1);x}(\Omega,\hat{\bm x}) = \chi^{(2j+1);y}(\Omega,\hat{\bm y})$ and  $\chi^{(2j+1);x}(\Omega,\hat{\bm y}) = \chi^{(2j+1);y}(\Omega,\hat{\bm x})=0$.
However, for even-order harmonics, which can only be generated in TQDs, their polarization is always along the $y$ direction with the response coefficients satisfying  $\chi^{(2j);y}(\Omega,\hat{\bm x}) = \chi^{(2j);y}(\Omega,\hat{\bm y})$ and $\chi^{(2j);x}(\Omega, \hat{\bm x}) = \chi^{(2j);x}(\Omega, \hat{\bm y})=0$.
All these results are consistent with the symmetry analysis. The TQDs have the $D_{3h}$ symmetry, which has no inversion center, and thus the even-order harmonics can be generated, and the perturbative second susceptibility components \cite{Boyd2008} satisfy $\chi^{(2);yyy} =-\chi^{(2);yxx}$, and the perturbative third-order susceptibility components satisfy $\chi^{(3);xxxx}=\chi^{(3);yyyy}$.
Note that the above discussions are also applicable to the results in other approximations (IPA, MFA, and RPA).
Next, we focus on the amplitude of the susceptibility $\chi^{(j);y}(\Omega,\hat{\bm y})$ for different GQDs.

\subsubsection{Susceptibility spectra at different approximations}

\begin{figure*}[!htb]
\centering
         \includegraphics[scale=0.9]{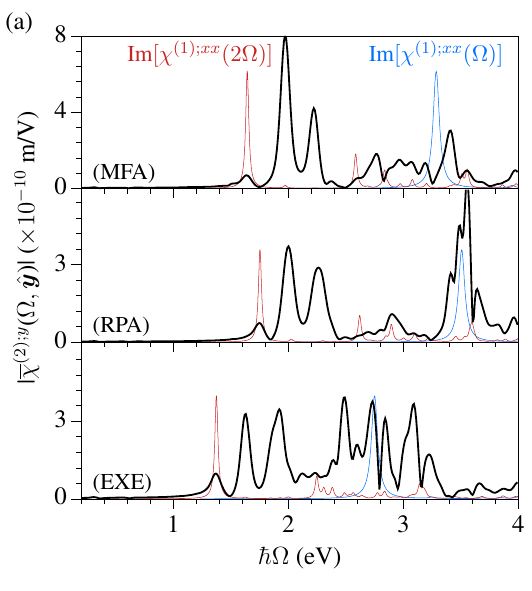}
         \includegraphics[scale=0.9]{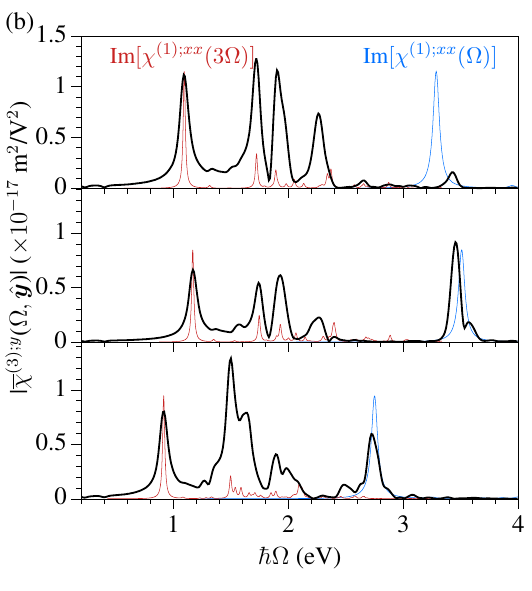}
         \includegraphics[scale=0.9]{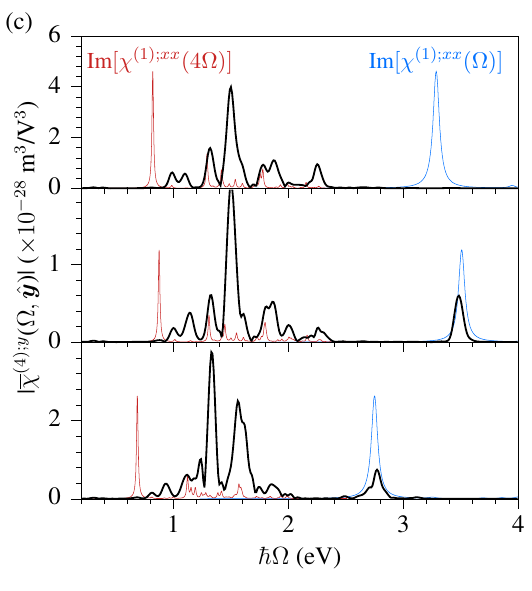}
         \includegraphics[scale=0.9]{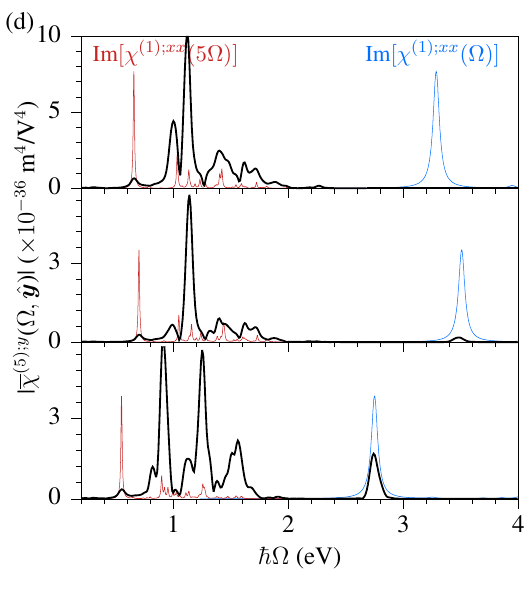}   				
      \caption{Photon energy $\hbar\Omega$ dependence of $\overline{\chi}^{(n);y}(\Omega,\hat{\bm y})$ for
      (a) SHG, (b) THG, (c) FOHG, and (d) FIHG of 4-AC-TQD under the approximations of MFA, RPA, and EXE. 
      The absorption spectra, $\text{Im}[\chi^{(1);xx}(\Omega)]$ (blue lines) and $\text{Im}[\chi^{(1);xx}(j\Omega)]$ (red lines) with $j=2$, 3, 4, or 5,  are plotted for each approximation.   
}
      \label{fig-5}
\end{figure*}

Now we investigate the effects of Coulomb interaction on the nonlinear optical responses of GQDs.
Figure\,\ref{fig-5} shows the spectra of $\overline{\chi}^{(j);y}(\Omega,\hat{\bm y})$ for SHG, THG, fourth harmonic generation (FOHG), and fifth harmonic generation (FIHG) of a 4-AC-TQD under approximations of MFA, RPA, and EXE  for a field amplitude of $10^8$ V/m.
Each figure also shows the linear absorption spectra 
$\text{Im}[\chi^{(1);xx}(\Omega)]$ to identify the photon energies for one-photon transition processes as well as $\text{Im}[\chi^{(1);xx}(j\Omega)]$ to identify the photon energies for $j$-photon transition processes.
All spectra show discrete peaks, which are consistent with transitions between discrete energy levels.
In the case of MFA,  the first peaks of spectra for $n$th-order harmonic are located at the same energy as the first peaks of $\text{Im}[\chi^{(1);xx}(j\omega)]$, which refers to the $n$-photon transition processes between the LUMO and HOMO states.
With the increase of photon energy, the optical transitions involve high energy states, and more peaks induced by multi-photon optical transitions appear because of the denser energy levels at higher energies, as shown in Fig.\,\ref{fig-2}. In our calculations, the relaxation energy is set to approximately 33 meV, and thus those peaks with similar transition energies merge into a broad peak. When the photon energy reaches  $\hbar\Omega\sim3.28$ eV, which is the energy for the first absorption peak, there should appear the first resonant peak for harmonic generation induced by the one-photon transition process \cite{Cheng2014}; however, in GQDs this peak is too small to be visible in the diagram, and the neighboring peaks are induced by the multi-photon resonant peaks. 
For the cases of RPA and EXE, the energy shift of all spectra is very similar to that of linear absorption, which shows the local plasmonic effects arising from the Hartree term and the EXE arising from the Fock term. 
Around the first absorption peak, which corresponds to $\hbar\Omega\sim3.5$ eV in RPA and $\hbar\Omega\sim2.7$ eV in EXE, the peak values of all harmonic spectra are notably small in MFA, but become well-pronounced in both RPA and EXE.
It is primarily attributed to the presence of local plasmonic effects around this energy range in the latter two, resulting in a significant enhancement of the effective electric field experienced by the electrons. 
These observations highlight the importance of considering many-body interactions and exciton effects in accurately describing the harmonic processes in GQDs.

\subsubsection{QD size dependence of the susceptibility spectra}
\label{sec-c-3}

\begin{figure*}[!htb]
         \includegraphics[scale=0.8]{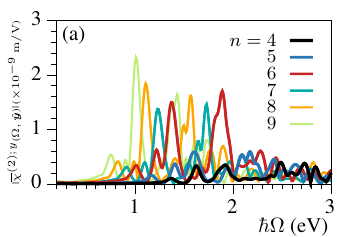}
         \includegraphics[scale=0.8]{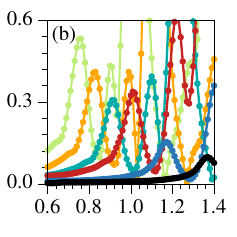}
         \includegraphics[scale=0.8]{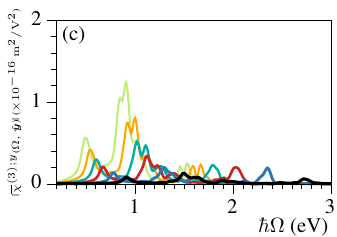}
         \includegraphics[scale=0.8]{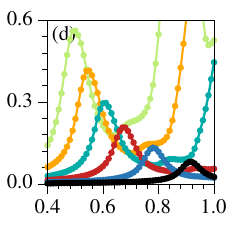}      
         \\
         \includegraphics[scale=0.8]{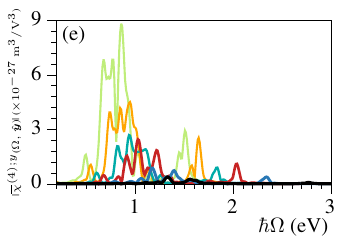}
         \includegraphics[scale=0.8]{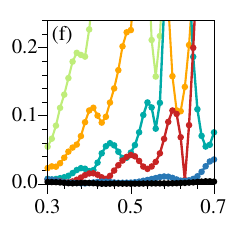}
         \includegraphics[scale=0.8]{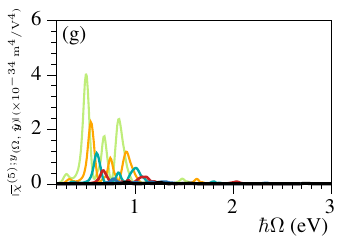}
         \includegraphics[scale=0.8]{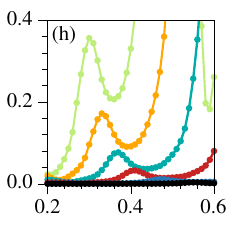}
         \caption{Spectra of nonlinear susceptibilities $|\bar{\chi}^{(j);y}(\Omega,\hat{\bm y})|$ of $N$-AC-TQD for different dot size under $E_0=10^8$ V/m with (a,b) $j=2$, (c,d) $j=3$, (e,f) $j=4$, and (g,h) $j=5$. For each harmonic order, the spectra around the first peak corresponding to the $j$-photon resonant transition are zoomed-in in the right figure. 
}
      \label{fig-6} 
\end{figure*}

Now we turn to the quantum dot size dependence of the nonlinear susceptibilities $\overline{\chi}^{(j);y}(\Omega,\hat{\bm y})$ of $N$-AC-TQD for $N$ from 4 to 9 in EXE under the field $E_0=10^8$ V/m, which are shown in Fig.\,\ref{fig-6}. 
It can be seen that these spectra show similar shapes with all characteristic peaks when the quantum dot size changes, however, both the locations and peak values vary with size.  
With increasing the QD size, the peak locations red-shift, similar to the absorption spectra in Fig.\,\ref{fig-3}. The reason is that the energy levels become denser, and thus the energy differences between energy levels decrease. 
The peak values show a complicated dependence on the quantum dot size. 
For example, in Fig.\,\ref{fig-6}\,(a), 
the peak value of the first peak, 
 corresponding to the two-photon resonant transition from the ground state to the lowest exciton state, approximately increases with $N$, except that the value at $N=6$ exceeds those at $N=5$ and $N=7$, see details in the zoomed-in Fig.\,\ref{fig-6}\,(b).
For other peaks, their peak values also increase with $N$.
Figures\,\ref{fig-6}\,(c--h) give the spectra of $\overline{\chi}^{(j);y}(\Omega,\hat{\bm y})$ for $j=3,4,5$ with different quantum dots. 
These spectra are similar to Figs.\,\ref{fig-6}\,(a) and (b).
For these three spectra, the peak values increase with the increase of quantum dot size. 
For GQDs with other shapes, all size-dependent properties are similar, except the first peak values of $\overline{\chi}^{(2);y}(\Omega,\hat{\bm y})$ reach a maximum at $N=6$, as shown in Fig.\,S1 in Supporting Information. 
In general, except for some special sizes, for large quantum dots, the peak locations are red-shifted to lower photon energies,  while the peak values increase.

For comparison, we note that 
the third-order nonlinear susceptibility $\chi^{(3)}$ of the monolayer graphene, measured via third harmonic generation,  ranges from 10$^{-19}$ to 10$^{-16}$ m$^2$/V$^2$ in experimental studies \cite{Hong2013,Kumar2013} and theoretical calculations \cite{Cox2014,Cheng2019}.
In this work, we find that 6-ZZ-HQD shows a maximum value of $\chi^{(3)}$  of 4.7$\times10^{-16}$ m$^2$/V$^2$ at $\hbar\Omega=0.9$ eV, representing comparable or superior nonlinear performance relative to monolayer graphene systems.
This  enhancement in nonlinear response is attributed to  the strong quantum confinements and edge effects at the nanoscale.


\subsubsection{Field intensity dependence}

\begin{figure}[!htb]
\centering
         \includegraphics[scale=0.8]{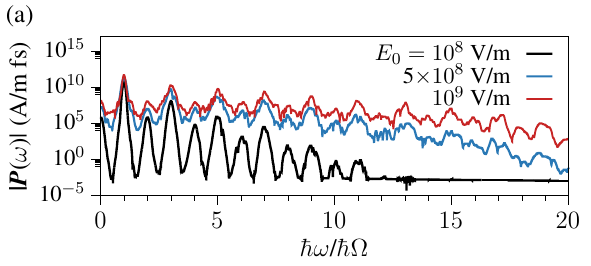}
         \includegraphics[scale=0.8]{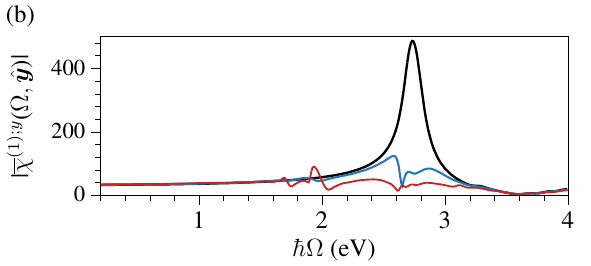}  
         \includegraphics[scale=0.8]{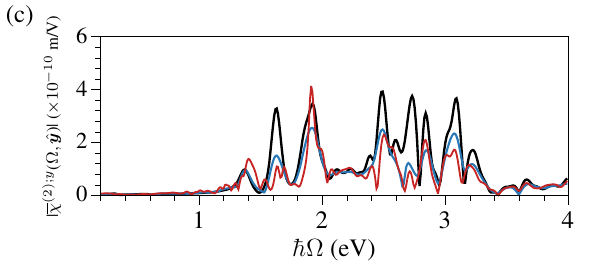}
         \includegraphics[scale=0.8]{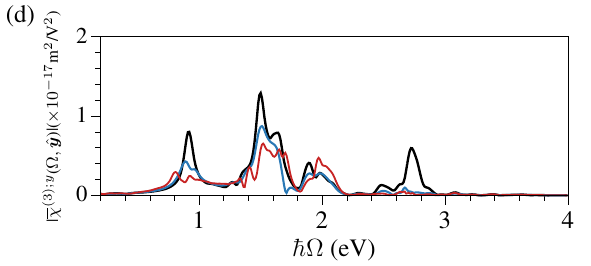}  
         \includegraphics[scale=0.8]{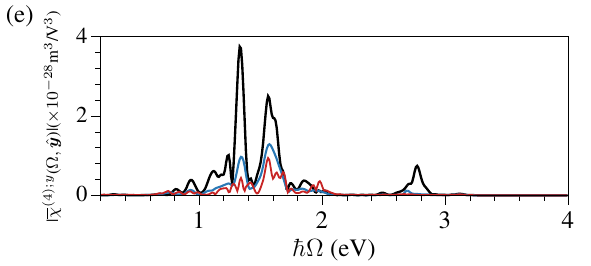}
         \includegraphics[scale=0.8]{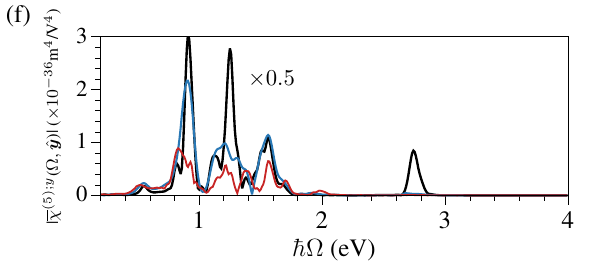}  
      \caption{(a) Spectra of $|{\bm P}(\omega)|$ with $\hbar\Omega=0.91$ eV and (b-f) spectra of $|\overline{\chi}^{(j);y}(\Omega,\hat{\bm y})|$ of the linear, SHG, THG, FOHG, and FIHG for 4-AC-TQD at different electric fields of $E_0=10^8$, $5\times10^8$, and $10^9$ V/m.
      Note the values of $|\overline{\chi}^{(5);y}(\Omega,\hat{\bm y})|$ with $E_0=10^8$ V/m, represented by the black line in (f), have been multiplied by 0.5.} 
      \label{fig-7}
\end{figure}

Figure\,\ref{fig-7}\,(a) shows the spectra of $|{\bm P}({\bm \omega})|$ of 4-AC-TQD for $\hbar\Omega=0.91$ eV under different electric fields, $E_0=10^8$, 5$\times10^8$, and $10^9$ V/m.
A stronger field strength leads to stronger polarization and higher order harmonics.
Figures\,\ref{fig-7}\,(b--f) give the  $|\overline{\chi}^{(j);y}(\Omega,\hat{\bm y})|$ for the linear, SHG, THG, FOHG, and FIHG under different electric fields.
For electric fields of $E_0=10^8$ and $5\times10^8$ V/m, the two spectra show minor changes in the peak locations, but significant reductions in peak amplitudes. 
This indicates that for these fields, harmonic responses can still be understood as optical transitions between exciton energy levels with  saturable absorption effects. In this case, the perturbation theory is applicable.
However, for stronger electric fields of  $E_0=10^9$ V/m, the spectra of $|\overline{\chi}^{(j);y}|$ are dramatically changed in both the peak energies and shapes, indicating a nonperturbative behavior. 
Note that higher order harmonics are more sensitive to the electric field strength.

\subsection{Polarization pattern dependence}

\begin{figure}[!htb]
\centering
         \includegraphics[scale=0.9]{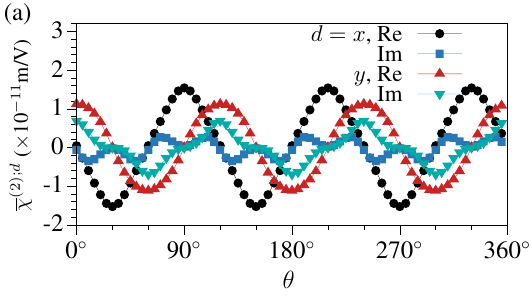}
         \includegraphics[scale=0.9]{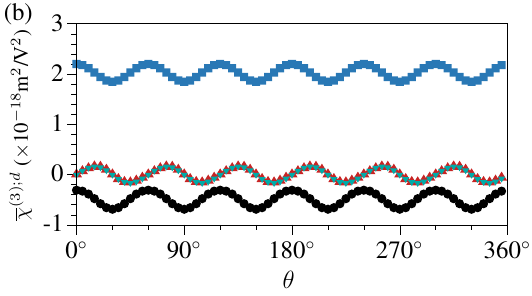}
         \includegraphics[scale=0.9]{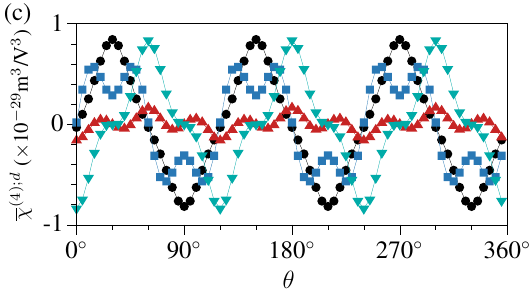}
         \includegraphics[scale=0.9]{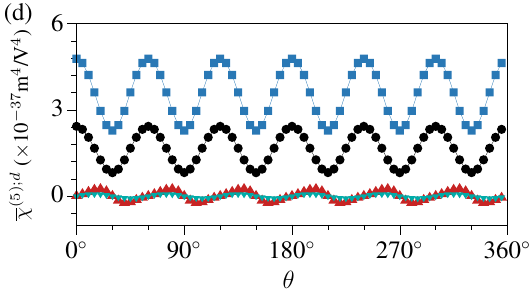}
      \caption{GQD orientation $\theta$ dependence  of $\overline{\chi}^{(j);d}$ for (a) SHG, (b) THG, (c) FOHG, and (d) FIHG in 4-AC-TQD at an  electric field $E_0=10^9$ V/m at $\hbar\Omega=0.91$ eV using the EXE approximation.}
      \label{fig-8}
\end{figure}

We now discuss the polarization pattern dependence of harmonic generation
for a GQD rotated about the $z$-direction by an azimuthal angle $\theta$, with incident light polarized along the $x$-direction. 
For very weak incident light, the polarization pattern of SHG and THG can be analyzed by considering the symmetry of each type of GQDs. 
For trigonal GQDs, as previously discussed, they have nonzero SHG susceptibilities due to the D$_{3h}$ symmetry.
For unrotated GQD with $\theta=0$, the nonzero susceptibility components for SHG are $\chi^{(2);xxy}=\chi^{(2);xyx}=\chi^{(2);yxx}=-\chi^{(2);yyy}=\chi^{(2)}$. 
Thus for a rotated GQD by angle $\theta$, the SHG polarization becomes  
\begin{align}
\bm P^{(2)}(\theta)=\chi^{(2)}E_x^2 (\hat{\bm x}\sin 3\theta - \hat{\bm y}\cos3\theta)\,.
\label{p2angle}
\end{align} 
In contrast, for hexagonal GQDs, SHG vanishes. 
For all GQD structures discussed in this work, the nonzero susceptibility components for THG are the same, and they are $\chi^{(3);xxxx} =\chi^{(3);yyyy}=\chi^{(3)}$ and $\chi^{(3);xxyy}=\chi^{(3);yyxx}=\chi^{(3);xyxy}=\chi^{(3);yxyx}=\chi^{(3);xyyx}=\chi^{(3);yxxy}=\chi^{(3)}/3$.
The resulting third harmonic polarization for a rotated GQD is always along the $x$-direction as 
\begin{align}
\bm P^{(3)}(\theta)=\chi^{(3)}E_x^3 \hat{\bm x} \,.
\label{p3angle}
\end{align} 
This reveals a fundamental
 symmetry distinction -- SHG exhibits a threefold (120$^\circ$) rotational periodicity, while THG remains isotropic and polarized along the same polarization direction as the incident light. 
All these results are confirmed by our numerical calculations for the incident electric field amplitude of $E_0=10^8$ V/m.  
Furthermore, higher even-order (odd-order) harmonics follow the angular dependence analogous to SHG (THG).
However, for stronger electric fields, the polarization pattern dependence deviates from 
the perturbative expressions in Eqs.\,\eqref{p2angle} and \eqref{p3angle}.
As shown in Fig.\,\ref{fig-8},
which illustrates the $\theta$-dependent susceptibilities for $E_0=10^9$ V/m and $\hbar\Omega=0.91$ eV,
the SHG and FOHG spectra  retain the threefold rotational periodicity.
Unlike the perturbative case, however, the amplitudes of $\overline{\chi}^{(j);x}$ and $\overline{\chi}^{(j);y}$ are no longer equal.
Additionally, while THG and FIHG retain predominantly $x$-direction polarization, they exhibit a near sixfold rotational periodicity.

Many experiments are performed on GQD ensembles formed by randomly orientated GQDs, and the measured polarization  is proportional to  $\frac{1}{2\pi}\int_0^{2\pi}\bm P^{(j)}(\theta)d\theta$. 
For very weak incident light,
utilizing Eqs.\,\eqref{p2angle} and \eqref{p3angle},  the SHG for GQD ensembles vanishes, while
the THG remains  unchanged.
In fact, this can also be understood through the symmetry consideration.
A GQD ensemble has isotropic symmety for which all even-order responses vanish due to the inversion symmetry, while all odd-order responses 
have the polarization direction aligned
with the external field.
For the nonperturbative case, the polarization of a GQD  ensemble is obtained  by numerically  integrating $\bm P^{(j)}(\theta)$  over $\theta$, yielding results similar to the perturbative case.
Consequently, for the macroscopic response in GQD ensembles, only odd-order harmonics
produce nonvanishing  isotropic polarization along the same direction as the incident light, consistent with results for in-plane isotropic systems.

\section{Conclusions}
\label{conclusions}

We present a comprehensive theoretical investigation of many-body effects on harmonic generations in graphene quantum dots, considering the influences of geometric parameters (shapes, sizes, and edges) and external electric fields.
The many-body interactions are modeled through a screened Hartree-Fock approximation and are incorporated into semiconductor Bloch equations. 
For the ground states, the antiferromagnetic state has the lowest energy, and the Coulomb interaction greatly affects the electronic energy levels and removes the spin degeneracy. 
For the optical responses, the many-body effects are studied under different approximations: the independent particle approximation, the mean-field approximation, the random phase approximation, and the excitonic effects.
The resonant transitions in the mean field approximation have larger transition energies than those in the independent particle approximation, indicating that the energy difference between the LUMO and HOMO levels is enlarged by Coulomb interaction, similar to the GW corrections; for the results obtained in the random phase approximation, the Hartree term induces local plasmonic resonance to enhance the electric field experienced by the electrons and further blue-shifts the resonant peaks in both linear and nonlinear optical responses; with considering the excitonic effects, the Fock term introduces attractive energy, which red-shifts the transition energies and greatly affects the resonant optical transitions.
Interestingly, although the $j$-photon process leads to a significant resonant peak for the $j$th-order harmonic generation, the one-photon process leads to negligible resonance in the mean field approximation; however, remarkable resonant peaks appear around the energy for the first linear absorption peak in the random phase approximation and the excitonic effects, indicating a strong influence of the Coulomb interaction on the harmonic generation.
We also discussed the impact of the geometric structure of quantum dots, including the size, shape, and edge shapes, on their nonlinear optical properties. 
Remarkably, the exciton states arising from these nano-structural features exhibit enhanced nonlinear susceptibility compared to that of monolayer graphene.
We also studied the electric field dependence.
For a very strong electric field, high order harmonics can be generated, but the susceptibility decreases; when the electric field intensity reaches $10^9$ V/m, the perturbation theory is no longer applicable.
Finally, the polarization pattern dependence is discussed for a single quantum dot and for an ensemble formed by randomly oriented graphene quantum dots. 
In the latter case,
all even-order harmonics vanish, while all odd-order harmonics exist along the polarization direction of the external field, which is similar to an in-plane isotropic system.
As a result, our analysis 
shows that graphene quantum dots exhibit size- and edge-tunable nonlinear optical responses,
which establishes a physical basis for 
possible applications in nonlinear optical devices.

\section{Methods}
\label{model}

%
%
By denoting the primitive lattice
vectors as ${\bm a}_1=(\sqrt{3}/2\hat{\bm x}-1/2\hat{\bm y})a_0$ and
${\bm a}_2=(\sqrt{3}/2\hat{\bm x}+1/2\hat{\bm y})a_0$ 
(see the inset of Fig.\,\ref{fig-1})
with a lattice
constant $a_0=2.46$ {\AA}, Table\,\ref{tab:1} lists the methods to
construct these GQDs as well as
their information about the numbers of hexagons ($N_h$) and carbon
atoms ($N_a$). The area $S_{\text{QD}}$ is then calculated as $S_{\text{QD}}=N_h\frac{\sqrt{3}}{2}a_0^2$.  
The position of each carbon atom can be
written as $\bm R_n=n_1{\bm a}_1+n_2\bm a_2+n_3(\bm a_1+\bm a_2)/3$ with
the abbreviated index $n=(n_1,n_2,n_3)$, where $n_1,n_2$ are integers
and $n_3=0$ ($1$) is for the A (B) site.

\begin{table*}[!htp]
  \centering
  \caption{Numbers of hexagons ($N_h$) and carbon
    atoms ($N_a$) for a $N$-QD, and methods to construct
      the centers of all hexagons whose vertexes can generate
      all locations of carbon atoms in the GQD
      (construction rules). 
      The construction gives the center
      positions as $\bm R_{c}+l_1\bm A_1 + l_2\bm A_2$, where $\bm
      R_c$ is a position bias that is not important, $\bm A_1$ and $\bm
      A_2$ are two vectors, $l_1$ and $l_2$ are integers with
      constraints.}
  {	
    \begin{tabular}{c|c|c|c|c}
      \hline\hline		
     GQD Types & $N_h$ & $N_a$ & $\bm A_i$ & Constraints\\
      \hline
      $N$-AC-TQD  &	$\frac{3}{2}(N^2-N)+1$	&	$3(N^2+N)$
      &\makecell[c]{$\bm A_1=\bm a_1+\bm a_2$,\\ $\bm A_2=2\bm a_2-\bm a_1$}&\multirow{2}{*}{$0\le l_1,l_2,l_1+l_2<N$}\\
      \cline{1-4}
      $N$-ZZ-TQD	&$\frac{1}{2}(N^2+N)$&
$N^2+4N+1$
      &$\bm A_i=\bm a_i$ &\\
      \hline
      $N$-AC-HQD	&$9N^2-15N+7$&$6(3N^2-3N+1)$
      &\makecell[c]{$\bm A_1=2\bm a_1-\bm a_2$,\\ $\bm A_2=2\bm a_2-\bm a_1$}
      &\multirow{3}{*}{$\begin{array}{c}0\le l_1,l_2<2N-1\\
                               |l_1-l_2|<N
                             \end{array}$ }\\
      \cline{1-4}
      $N$-ZZ-HQD&$3N^2-3N+1$&$6N^2$
      &\makecell[c]{$\bm A_1=\bm a_1$,\\ $\bm A_2=\bm a_2-\bm a_1$} &\\
      \hline\hline
    \end{tabular}}
  \label{tab:1}
\end{table*}%

In a simple tight-binding model formed by the $2p_z$ carbon orbitals, the
Hamiltonian is 
\begin{align}
\label{h-main}
    \hat{H}(t) =& \sum_{nm, \sigma} h_{nm} \hat{a}_{n\sigma}^\dag(t)\hat{a}_{m\sigma}(t) \notag \\
    &+\frac{1}{2}\sum_{nm,\sigma\sigma'}V_{nm}\hat{a}_{n\sigma}^\dag(t)\hat{a}_{m\sigma'}^\dag(t)\hat{a}_{m\sigma'}(t)\hat{a}_{n\sigma}(t) \notag \\
	&+|e|{\bm E}(t)\cdot\sum_{n,\sigma}{\bm R}_{n}\hat{a}_{n\sigma}^\dag(t)\hat{a}_{n\sigma}(t)\,,
\end{align}
where $\hat{a}_{n\sigma}(t)$ ($\hat{a}_{n\sigma}^\dag(t)$) is the
annihilation (creation) operator for the orbital with spin
  $\sigma$ located at ${\bm R}_n$, 
$h_{nm}$ gives the hopping energy
  between the sites $n$ and $m$ where only the nearest neighbour
  coupling  \cite{Cheng2015a,Cheng2014,Cheng2014a,Cheng2019} is considered with a coupling energy $\gamma_0=-2.7$ eV, $e$ is the electron charge, and ${\bm E}(t)$ is the applied electric field.
The second term on the  right-hand side of Eq.\,\eqref{h-main}
gives the electron-electron interaction and $V_{nm}$ is the Ohno potential \cite{Sadeq2018,Jiang2007} as
\begin{align}
V_{nm}=\frac{U}{\epsilon\sqrt{\left(\frac{4\pi\epsilon_0|{\bm R}_n-{\bm R}_m|U}{e^2}\right)^2+1}}\,,
\label{ohno}
\end{align} 
where $U=8.29$ eV is the onsite energy and $\epsilon=3.5$ is used to describe the effective background dielectric constant. 
%


With neglecting the spin-orbit coupling, the electron states can
be described by the density matrix
\begin{align}
\label{rho}
\rho_{nm;\sigma}(t)= \langle \hat{a}_{m\sigma}^\dag (t) \hat{a}_{n\sigma}(t) \rangle\,,
\end{align}
and their dynamics can be described by the semiconductor Bloch
equations (SBE) derived from the Green function method at the
  level of static screened HF approximation
\begin{align}
\label{sbe}
i \hbar \frac{\partial \rho_{nm;\sigma} (t)}{\partial t} = [\mathcal{H}_\sigma(t), \rho_\sigma(t)]_{nm} 
- i\Gamma[\rho_{nm;\sigma}(t)-\rho_{nm;\sigma}^0] \,.
\end{align}
The last term describes the phenomenological relaxation
described by the damping parameter $\Gamma$,
and $\rho_{nm;\sigma}^0$ gives the density matrix
  at the ground state. The elements of the matrix $\mathcal{H}_\sigma(t)$ are 
\begin{align}
[\mathcal{H}_\sigma(t)]_{nm}
= h_{nm}+\delta_{n m}\left[|e| {\bm E}(t) \cdot {\bm R}_{n}+
  \sum_{l,\sigma^\prime} V_{n l} \rho_{ll;\sigma^\prime}(t) \right] - \mathcal{W}_{n m}^{0} \rho_{n m;\sigma}(t)\,.
\label{h-matrix}
\end{align}
Here the term involving $V$
  gives the Hartree contribution and the last term gives the Fock
  contribution \cite{Cheng2019} where $\mathcal{W}_{n m}^{0}=W_{nm}^0(0)$ is
  the static screened Coulomb interaction at equilibrium states. 
  The Hartree term describes the Coulomb potential from the
  inhomogeneous local charge density, which also induces the local
  field effects during the optical responses; while the Fock
  term includes the EXE. The method to determine $\rho^0$ and ${\cal W}^0$ is given below.

  When  there is no external electric field,  the unperturbed 
  Hamiltonian can be written as
  \begin{subequations}
  \begin{align}
    {\cal H}_{nm;\sigma}^0
    &=h_{nm}+\delta_{n m}\lambda_{0h}
      \sum_{l,\sigma^\prime} V_{n l} \rho_{ll,\sigma^\prime}^0-
      \lambda_{0f}\mathcal{W}_{n m}^{0} \rho_{n m;\sigma}^0\,,\label{eq:H0}
  \end{align}
  where $\lambda_{0h}$ and $\lambda_{0f}$ are parameters introduced to turn on
    ($=1$) or off  ($=0$) the contributions from Hartree and Fock terms at
    the ground state.
    The electronic eigenstates are obtained from the Schr\"odinger
  equation
  \begin{align}
    \sum_m {\cal H}_{nm;\sigma}^0 C_{s;m\sigma} &= \varepsilon_{s\sigma}
                                                  C_{s\sigma;n}\,,\label{eq:sch}
  \end{align}
  where $s$ is the band index for electronic levels. The 
  density matrix at the  ground state is
  \begin{align}  
  \rho_{nm;\sigma}^0 &= \sum_s f_{s\sigma} C_{s\sigma;n}C_{s\sigma;m}^\ast\,,\label{eq:rho0}
  \end{align}
  where $f_{s\sigma}=[1+e^{-(\varepsilon_{s\sigma}-\mu)/(k_BT)}]^{-1}$ is the
Fermi-Dirac distribution at a temperature $T$ and chemical potential
$\mu$. 
The chemical potential is related to the total electron
  number $N_e$ 
  of the GQD through $\sum_{s\sigma}f_{s\sigma}=N_e$.
Thus the Green function at the ground state is
\begin{align}
    G_{n m;\sigma}^{0}(\omega)&=
\sum_{s}\left[\frac{1-f_{s\sigma}}{h \omega-\varepsilon_{s\sigma}+i
                                0^{+}}+\frac{f_{s\sigma }}{h
                                \omega-\varepsilon_{s\sigma }-i
                                0^{+}}\right] C_{s\sigma;n}
                                C_{s\sigma;m}^{*}\,,\label{eq:g0}
\end{align}
from which the polarization is
\begin{align}
    P_{n m}^{0}(\omega)&=-i \hbar \sum_\sigma \int \frac{d \omega_{1}}{2 \pi} G_{n
                         m;\sigma}^{0}\left(\omega_{1}-\omega\right) G_{m
                         n;\sigma}^{0}\left(\omega_{1}\right)\,.\label{eq:p0}
\end{align}
Then the screened Coulomb interaction is written in matrix form as
\begin{align}
    W^0(\omega) &= \left[V^{-1}-P^0(\omega)\right]^{-1}\,.\label{eq:w0}
  \end{align}
\end{subequations}
Equations\,\eqref{eq:H0}--\eqref{eq:w0} are coupled and they can be solved iteratively. 
For graphene nanostructures with different edges, 
the ground states can be magnetic \cite{Yazyev2010}. To consider the possibilities of
magnetic ground states, the iteration is performed as follows: 
starting with a Hamiltonian matrix ${\cal H}_{nm;\sigma}^0=h_{nm}+\delta_{nm}M_{n;\sigma}$ where $M_{n;\sigma}$ is an
artificial onsite term depending on the initial magnetic configuration, we
iteratively solve
Eqs.\,\eqref{eq:rho0}--$\cdots$--\eqref{eq:w0}--\eqref{eq:H0}--\eqref{eq:sch}--$\cdots$
until achieving the convergence of the density matrix $\rho^0$. The
ground state is chosen as the lowest energy state among  the
paramagnetic state using $M_{n;\sigma}=0$, the ferromagnetic state using
$M_{n;\sigma}=\sigma \Delta$, and  the antiferromagnetic state using
$M_{n;\sigma}=\sigma (2n-1) \Delta$ with $\Delta=0.1$ eV.

By applying the electric field ${\bm E}(t)$, the dynamics of the density
matrix is obtained by solving the SBE in Eq.\,\eqref{sbe}
numerically. 
To better understand the local field effects and the EXE,
  we separate the density matrix into two terms
\begin{align}
\rho(t)=\rho^0+ \rho^e(t)\,. 
\end{align}
Then the Hamiltonian in Eq.\,\eqref{h-matrix} can be rewritten as
$\mathcal{H}(t)=\mathcal{H}^0 + \mathcal{H}^e(t) $ with
\begin{align}
\label{H-hf}
  \mathcal{H}_{nm;\sigma}^e(t)
  =& \delta_{n m}\left[\lambda_{h}  \sum_{l,\sigma'} V_{nl} \rho_{ll; \sigma'}^e(t)+ |e| \boldsymbol{E}(t) \cdot  \boldsymbol{R}_{n}\right]
     -\lambda_{f} \mathcal{W}_{nm}^{0}
     \rho_{nm\sigma'}^{e}(t)\,,
\end{align}
where $\lambda_h$ and $\lambda_f$ are parameters introduced to turn on
($=1$) or off  ($=0$) the contributions from Hartree and Fock terms
induced by the excited density matrix $\rho^e$.

In this work, we are interested in the sheet polarization
density  $\bm P(t)$ of the
system, which is defined as
\begin{align}
{\bm P}(t)=\frac{-|e|}{S_{\text{QD}}d_{\rm gr}} \sum_{n} {\bm R}_{n} \sum_\sigma\rho_{nn;\sigma}^{e}(t)\,,\label{eq:pt}
\end{align}
with the effective thickness of graphene $d_{\rm gr}=3.3$ {\AA}.
The definition in Eq.\,\eqref{eq:pt} excludes the static
  polarization, thus the absolute coordinate of the gravity center of
  the GQD does not affect $\bm P(t)$. For weak electric fields, the
  perturbative susceptibilities $\chi^{(i)}$ with different orders can be defined through
  \begin{align}
    P^d(t) =& \int \frac{d\omega}{2\pi}
    \chi^{(1);da}(\omega)E^a(\omega) e^{-i\omega t} \notag\\
    &+  \int \frac{d\omega_1d\omega_2}{(2\pi)^2}
    \chi^{(2);dab}(\omega_1,\omega_2)E^a(\omega_1)E^b(\omega_2)
    e^{-i(\omega_1+\omega_2) t} \notag\\
    &+ \int \frac{d\omega_1d\omega_2d\omega_3}{(2\pi)^3}
    \chi^{(3);dabc}(\omega_1,\omega_2,\omega_3)E^a(\omega_1)E^b(\omega_2) E^c(\omega_3)e^{-i(\omega_1+\omega_2+\omega_3) t} \notag\\
    &+\cdots\,,
  \end{align}
  where $\bm E(\omega)=\int dt \bm E(t) e^{i\omega t}$ gives the
  Fourier transform of $\bm E(t)$. 

\begin{table}[!htp]
	\centering
	\caption{Approximations used in this work.}
	{	
	\begin{tabular}{ccc|ccc}
		\hline\hline		
Ground States & $\lambda_{0h}$ & $\lambda_{0f}$ & $\lambda_{h}$ & $\lambda_{f}$ & Response \\
	    \hline
IPA  &	0	&	0	&	0	&	0	& IPA\\
		\hline
\multirow{3}{*}{MFA}	 &	1	&	1	&	0	&	0	& MFA\\
	 &	1	&	1	&	1	&	0	& RPA\\
	 &	1	&	1	&	1	&	1	& EXE\\
	    \hline\hline
	\end{tabular}}
      \label{tab:app}
\end{table}  

  Before we present the results, we give a brief summary of the approximations with the four artificially introduced parameters $\lambda_{0h}$,
  $\lambda_{0f}$, $\lambda_h$, and $\lambda_f$. 
  For the ground states, the absence of the electron-electron interaction
  with $\lambda_{0h}=\lambda_{0f}=0$ leads to IPA based on the tight-binding model, while its
  presence with $\lambda_{0h}=\lambda_{0f}=1$ gives the ground states
  at HF approximation, which is usually called MFA. 
  For the optical response,  $\lambda_{h}=1$ includes the feedback
  of the optical excited charge density distribution, giving the
  widely discussed RPA \cite{Cox2015}; while
  a combination of $\lambda_{h}=1$  and $\lambda_f=1$ refers to EXE including the exchange-correlation effects for the optically excited electron-hole pairs, which was not well studied in the literature and is
  the focus of this work. In Table\,\ref{tab:app} we list all used
  approximations.

\begin{acknowledgement}

This work has been supported by National Natural Science Foundation of China Grant No. 12034003.

\end{acknowledgement}

\begin{suppinfo}

Spectra of nonlinear susceptibilities for different sizes of triangular graphene quantum dots with zigzag edges and hexagonal graphene quantum dots with armchair and zigzag edges (PDF).

\end{suppinfo}
\bibliography{ref}

\end{document}